\documentclass[12pt]{article}
\usepackage{amssymb,amsmath,epsfig}
\allowdisplaybreaks

\begin{document}
\title{\textbf{Investigating Strange Stars in Rastall Theory}}
\author{Malick Sallah$^{1,2}$ \thanks{malick.sallah@utg.edu.gm}~and
M. Sharif$^1$ \thanks{msharif.math@pu.edu.pk (Corresponding author)}  \\
$^1$ Department of Mathematics and Statistics, The University of Lahore\\
1-KM Defence Road Lahore-54000, Pakistan.\\
$^2$ Department of Mathematics, The University of The Gambia,\\
Serrekunda, P.O. Box 3530, The Gambia.}

\date{}
\maketitle
\begin{abstract}
This study explores the structural formation of various spherically
symmetric anisotropic stars within the framework of Rastall theory.
To achieve this, we derive modified field equations which are then
resolved using the Finch-Skea ansatz, which involve unknown
parameters $(A_1,A_2,A_3)$. These parameters are found by using
appropriate constraints given by the junction conditions, in
addition to observational data from some selected stars. The
$\mathbb{EOS}$ given by the $\mathbb{MIT}$ bag model is employed to
examine the interior structure and various physical properties of
these compact objects. For calculated values of the bag constant
$\mathcal{B}$ and two values of the Rastall parameter, $\xi = 0.3,
0.5$, we investigate the regularity and viability of the state
variables. Additionally, we analyze stability of the developed model
by employing three distinct criteria. We find that the obtained
model is stable and provides an accurate approximation for the mass
and radius of strange stars when the Rastall parameter $\xi = 0.3$
is considered.
\end{abstract}

\section{Introduction}

General relativity (GR) is thought by many researchers to extend on
a cosmological scale as it is not thoroughly tested in environments
of extreme gravitational strength, such as those found near black
holes. Furthermore, GR does not readily account for the observed
rapid increase in the expansion of the cosmos unless the presence of
hypothetical entities like dark matter and energy are assumed. There
are two main strategies for modifying GR: one involves preserving
its core principles while introducing new terms in the Lagrangian
density, resulting in altered field equations; the other approach
involves changing some of GR foundational assumptions. Rastall
theory of gravity belongs to the latter group. Rastall \cite{1}
argued that the zero divergence of the Einstein tensor does not
automatically imply that the energy-momentum tensor also has zero
divergence.

The Rastall theory has faced criticism \cite{1a,1aa}, particularly
regarding the lack of conservation of energy-momentum tensor, a
claim contested by other authors \cite{1b,1c}. However, this
perceived violation can be seen as a consequence of spacetime
curvature or even the net creation of energy in certain systems.
Another common criticism is the absence of a Lagrangian formulation
for the theory, despite its success in producing acceptable results
in both cosmology and astrophysics. Attempts to derive a suitable
Lagrangian have been unsuccessful so far, raising doubts about its
feasibility. Despite these challenges, the advantages of the Rastall
theory are notable, with various theoretical and observational
studies appearing in recent research \cite{2}-\cite{5}. In more
recent studies, the role of the Rastall parameter in constructing
novel stellar solutions within spherical symmetry \cite{5a} as well
as in various models involving complexity and isotropization
\cite{5aa,5aaa} has been investigated. Stability analysis of
anisotropic stellar structures has also been studied in this theory,
using the cracking technique \cite{5b}. Waseem and Naeem \cite{5bb}
employed the Durgapal-Lake solutions to study isotropic stellar
models in this theory. We have also profited from this theory to
obtain spherically symmetric anisotropic solutions \cite{5c,5cc} as
well as extended black hole solutions \cite{5d}-\cite{5e} in this
theory.

The development and transition of cosmic structures throughout the
cosmos is significantly influenced by stars. Over the years, many
astrophysicists have dedicated their research to the investigation
of the evolution and interior geometry of celestial structures. The
inward gravitational pull due to the mass of a stellar object is
countered by an opposing push resulting from nuclear reactions
taking place in its core. However, when this pressure can no longer
counteract gravity, the star undergoes a gravitational collapse,
leading to its end and giving rise to different remnants, known as
compact objects. Owing to their intriguing composition and geometric
features, neutron stars have gained the attention of many
astrophysicists and researchers. In these stars, the balance between
neutron degeneracy pressure and gravitational force sustains
hydrostatic equilibrium. A quark star is an ultra-dense celestial
object that that is denser than a neutron star but less dense than a
black hole.

Currently, the investigation of compact objects with interiors
containing anisotropic matter has become a compelling topic of study
for many astronomers. Herrera \cite{12} noted that celestial bodies
with a nuclear density at their core significantly lower than their
mass density should be characterized by anisotropic fluids. Kalam et
al. \cite{13} formulated solutions for field equations corresponding
to various neutron stars, demonstrating their stability and
viability. Effective solutions for compact stars in hydrostatic
equilibrium were provided by Paul and Deb \cite{14}. Within the
framework of Rastall theory, Tangphati et al. \cite{15} examined the
interior geometry and physical properties of quark stars, while
Salako et al. \cite{16} investigated how electromagnetic fields
influence strange quark matter within a quintessence field .
Panotopoulos et al. \cite{17} conducted a detailed examination of
strange quark star matter under Lovelock gravity and standard
theory, assuming pressure anisotropy. Bhar \cite{18} developed
another anisotropic model for strange stars.

Mota et al. \cite{23a} generalized the Rastall gravity framework to
derive the field equations for spherically symmetric compact
objects, demonstrating its impact on neutron star structures.
Similarly, Nashed and Hanafy \cite{23aa} investigated anisotropic
compact stars within the Rastall framework, showing that the Rastall
parameter significantly influences the physical properties of these
stars, such as their mass-radius relations and stability conditions.
Additionally, El Hanafy \cite{23b} applied Rastall gravity to model
the pulsar PSR $J0740+6620$, finding that the model aligns well with
observational constraints and provides valuable insights into the
mass and radius of compact stellar objects. In nonminimally coupled
gravity, Sharif and Naseer \cite{20}-\cite{23} studied various
anisotropic strange stars.

The $\mathbb{MIT}$ bag model equation of state ($\mathbb{EOS}$) is
expected to provide a means of describing the internal structure of
quark stars \cite{24}. This model is especially useful in explaining
the compactness of certain astronomical objects, such as \emph{RXJ}
$\emph{185635-3754}$, \emph{4U} $\emph{1728-34}$, \emph{Her}, etc.,
which are not accountable via $\mathbb{EOS}$ for neutron stars. The
compactness of these objects is effectively explained by the
$\mathbb{MIT}$ bag model \cite{25}. The difference between true and
false vacuum states can be determined by the bag constant
$\mathcal{B}$ in the bag model $\mathbb{EOS}$, with increments in
this constant resulting in a decrease in quark pressure. Several
researchers \cite{26,27} have utilized this $\mathbb{EOS}$ to
predict interior dynamics of quark stars. By measuring the mass of
\emph{PSR} $\emph{J1614-2230}$, it was found that only this
$\mathbb{EOS}$ could account for such a massive object \cite{28}. By
analyzing physical features of a star with a radius of
$9.9~\text{km}$, the masses of various stars were calculated by
using an interpolating function \cite{29}.

Deb et al. \cite{32,33} focused on both charged and uncharged
strange stars, developing regular solutions based on this
$\mathbb{EOS}$ and validated their results through graphical
analysis. Sharif et al. \cite{34,35} extended this work by deriving
anisotropic solutions for various stellar candidates using the
$\mathbb{MIT}$ bag model. Celestial bodies with masses ranging from
$8$ to $20$ times that of the Sun collapse to form neutron stars.
Depending on their densities, these neutron stars can further evolve
into black holes or quark stars \cite{35a}. Notably, these stars are
characterized by strong gravitational fields due to their highly
dense nature despite their small size.

Driven by these aforementioned studies, this paper studies the
potential existence of strange compact stars within the framework of
Rastall gravity. The study examines distinct physical features of
the proposed model using experimental data from five known compact
stars, deriving quantitative results for relevant physical
variables. The outline of this paper is as follows. In section
\textbf{2}, we present the derivation of the field equations in
Rastall gravity and their corresponding solutions. In section
\textbf{3}, we determine the quantitative results of the parameters
arising from the Finch-Skea ansatz by matching the outer and inner
geometries. Section \textbf{4} provides a detailed graphical
analysis of the physical properties of the results. Finally, section
\textbf{5} offers a summary and conclusion of the obtained results.

\section{Formalizing the Rastall Field Equations}

In Rastall theory, the field equations deviate from those of GR due
to the presence of the Rastall parameter $(\zeta)$, which links the
covariant divergence of the Rastall stress-energy tensor to the
divergence of the curvature scalar $(\mathcal{R})$, as expressed
below
\begin{equation}\label{1}
\nabla^\chi T_{\eta\chi}=\zeta g_{\eta\chi}\nabla^\chi\mathcal{R}.
\end{equation}
Building on this concept, Rastall reinterpreted the Einstein field
equations by introducing a nonminimal interaction between geometry
and matter, formulated as follows \cite{1}
\begin{equation}\label{2}
\mathcal{R}_{\eta\chi}-\frac{1}{2}\mathcal{R}g_{\eta\chi}
+\xi\mathcal{R}g_{\eta\chi}=\kappa\tilde{T}_{\eta\chi},
\end{equation}
where $\xi=\kappa\zeta$ denotes the Rastall dimensionless parameter.
The aforementioned field equations reduce to those of GR when $(\xi
= 0)$. Furthermore, $( \tilde{T}_{\eta\chi})$ represents an
energy-momentum tensor associated with an anisotropic matter
configuration, expressed as follows
\begin{equation}\label{3}
\tilde{T}_{\eta\chi}=(\tilde{P}_r-\tilde{P}_t)\mathcal{Z}_\eta
\mathcal{Z}_\chi-\tilde{P}_tg_{\eta\chi}+(\tilde{\rho}+\tilde{P}_t)
\mathcal{W}_\eta\mathcal{W}_\chi.
\end{equation}
Here,
\begin{itemize}
\item $\mathcal{W}$ is the 4-vector,
\item $\mathcal{Z}$ is the 4-velocity,
\item $\tilde{\rho}$ is the density,
\item $\tilde{P}_r$ is the radial pressure,
\item $\tilde{P}_t$ is the tangential pressure.
\end{itemize}
Furthermore,
\begin{equation}\label{4}
\mathcal{W}^\eta=\delta_0^\eta\sqrt{g^{00}},\quad
\mathcal{Z}^\eta=\delta_1^\eta\sqrt{-g^{11}},
\end{equation}
satisfy the relations $\mathcal{W}^\eta\mathcal{W}_\eta=1$,
$\mathcal{W}^\eta\mathcal{Z}_\eta=0$, and
$\mathcal{Z}^\eta\mathcal{Z}_\eta=-1$.

From the field equations \eqref{2}, we derive
\begin{equation}\label{5}
\mathcal{R}(4\xi-1)=\kappa \tilde{T},
\end{equation}
depicting the invalidity of $\xi=\frac{1}{4}$. As $\xi=\kappa\zeta$
in the Newtonian limit, then
\begin{equation}\label{6}
\kappa=\frac{(4\xi-1)8\pi}{6\xi-1},
\end{equation}
and
\begin{equation}\label{7}
\zeta=\frac{(6\xi-1)\xi}{(4\xi-1)8\pi}.
\end{equation}
Using Eq.\eqref{6}, the field equations become
\begin{equation}\label{8}
\mathcal{R}_{\eta\chi}-\frac{1}{2}\mathcal{R}g_{\eta\chi}
+\xi\mathcal{R}g_{\eta\chi}=\frac{4\xi-1}{6\xi-1}8\pi\tilde{T}_{\eta\chi},
\end{equation}
following which
\begin{equation}\label{9}
\mathcal{R}(6\xi-1)=8\pi\tilde{T}.
\end{equation}
This shows that $\xi=\frac{1}{6}$ is also inadmissible in the
Rastall theory. Furthermore, the metric
\begin{equation}\label{10}
ds^2_-=e^{\mu(r)}dt^2-e^{\nu(r)}dr^2-r^2(d\theta^2+\sin^2\theta
d\phi^2),
\end{equation}
is used to denote the interior geometry. With this metric, the field
equations \eqref{8} are obtained as
\begin{align}\nonumber
\left(\frac{4\xi-1}{6\xi-1}\right)8\pi\tilde{\rho}&=\xi\bigg[e^{-\nu}\bigg(\mu^{\prime\prime}
+\frac{\mu^\prime}{2}(\mu^\prime-\nu^\prime)+\frac{2}{r}(\mu^\prime-\nu^\prime)
+\frac{2}{r^2}\bigg)-\frac{2}{r^2}\bigg]\\\label{11}&+e^{-\nu}\bigg(\frac{\nu^\prime}{r}
-\frac{1}{r^2}\bigg)+\frac{1}{r^2},
\end{align}
\begin{align}\nonumber
\left(\frac{4\xi-1}{6\xi-1}\right)8\pi
\tilde{P}_r&=-\xi\bigg[e^{-\nu}\bigg(\mu^{\prime\prime}
+\frac{\mu^\prime}{2}(\mu^\prime-\nu^\prime)+\frac{2}{r}(\mu^\prime-\nu^\prime)
+\frac{2}{r^2}\bigg)-\frac{2}{r^2}\bigg]\\\label{12}&+e^{-\nu}\bigg(\frac{\mu^\prime}{r}
+\frac{1}{r^2}\bigg)-\frac{1}{r^2},
\end{align}
\begin{align}\nonumber
\left(\frac{4\xi-1}{6\xi-1}\right)8\pi\tilde{P}_t&=
-\xi\bigg[e^{-\nu}\bigg(\mu^{\prime\prime}+\frac{\mu^\prime}{2}(\mu^\prime-\nu^\prime)
+\frac{2}{r}(\mu^\prime-\nu^\prime)+\frac{2}{r^2}\bigg)-\frac{2}{r^2}\bigg]\\\label{13}
&+e^{-\nu}\bigg[\frac{\mu^{\prime\prime}}{2}+\frac{\mu^{\prime^2}}{4}
-\frac{\mu^\prime\nu^\prime}{4}+\frac{\mu^\prime-\nu^\prime}{2r}\bigg].
\end{align}
The system above comprises three equations in the five unknowns:
$\tilde{\rho},\mu,\nu,\tilde{P}_r$, $\tilde{P}_t$. In order to study
the interior structure of quark stellar configurations in the
context of the Rastall theory, the field equations above must be
explored together with the well-known $\mathbb{MIT}$ bag model
$\mathbb{EOS}$. This model \cite{36,37}
\begin{equation}\label{14}
\tilde{P}_r-\frac{1}{3}(\tilde{\rho}-4\mathcal{B})=0,
\end{equation}
where $\mathcal{B}$ denotes the bag constant, establishes a
relationship between state parameters of compact stellar structures
and plays a crucial role in defining the unique features of quark
stars. As highlighted in the literature, this model has widely been
used by many researchers to study the internal distribution of these
stars.

We thus consider the system \eqref{11}-\eqref{14}  of four
equations, comprising of six unknowns:
$\tilde{\rho},\mu,\nu,\tilde{P}_r,\tilde{P}_t, \mathcal{B}$. It thus
turns out that two additional constraints are required to balance
this system. For this purpose, we employ the ansatz given by
Finch-Skea spacetime \cite{39}
\begin{equation}\label{15}
e^{\nu(r)}=A_3r^2+1,\quad
e^{\mu(r)}=\big(A_1+\frac{1}{2}A_2r\sqrt{A_3r^2}\big)^2,
\end{equation}
where $A_1,A_2,A_3$ are parameters to be determined using the
matching conditions. The Finch-Skea ansatz is chosen because it
allows for analytical tractability while providing a robust
foundation for studying the impact of anisotropic pressures and the
Rastall parameter on the structural properties of compact stars.
Moreover, the Finch-Skea ansatz has demonstrated success in previous
studies of compact stars, including those modeled using modified
theories of gravity, which strengthens its applicability in the
present context. Notably, it has been explored in Rastall gravity by
Sharif and Sallah \cite{5c,5cc}, who investigated anisotropic
stellar structures using this metric and found that it effectively
captures the influence of the Rastall parameter on compact stars.
Additionally, recent studies such as Shahzad et al. \cite{39a} and
Mustafa et al. \cite{39aa} have employed the ansatz to construct
stable anisotropic stellar models, while Bhar et al. \cite{39bb}
explored its effectiveness in describing quark stars under
alternative gravity frameworks. Furthermore, Sharif and Manzoor
\cite{39b} utilized the Finch-Skea ansatz to analyze equilibrium and
stability conditions in gravitational decoupling scenarios.

From the system given by Eqs.\eqref{11}-\eqref{14}, we derive the
expressions below
\begin{align}\label{16}
\tilde{\rho}&=\frac{(1-6\xi) \left(-\frac{3 A_3}{\left(A_3
r^2+1\right){}^2}-\frac{6 A_2 \sqrt{A_3 r^2}}{r \left(A_3
r^2+1\right) \left(A_2 r \sqrt{A_3 r^2}+2 A_1\right)}+\frac{16
\pi\mathcal{B} (1-4 \xi )}{6 \xi -1}\right)}{16 \pi  (4 \xi
-1)},\\\label{17}\tilde{P}_r&=\frac{(6 \xi -1) \left(A_2 \sqrt{A_3
r^2} \left(3 A_3 r^2+2\right)+2 A_1 A_3 r\right)}{16 \pi  (4 \xi -1)
r \left(A_3 r^2+1\right){}^2 \left(A_2 r \sqrt{A_3 r^2}+2
A_1\right)}-\mathcal{B},\\\nonumber\tilde{P}_t&=\frac{1}{16 \pi (4
\xi -1) \sqrt{A_3 r^2} \left(A_3 r^2+1\right){}^2 \left(A_2 r
\sqrt{A_3 r^2}+2 A_1\right)}\bigg[A_2 A_3 r \bigg[A_3
r^2\\\nonumber&\times \bigg[2 A_3 (4 \xi -1) r^2 \left(8 \pi
\mathcal{B} r^2+6 \xi -1\right)+32 \pi \mathcal{B} (4 \xi -1) r^2-48
\xi ^2+38 \xi\\\nonumber& -5\bigg]+2 \left(8\pi\mathcal{B} (4 \xi
-1) r^2+6 \xi (11-24 \xi )-7\right)\bigg]+2 A_1 \sqrt{A_3 r^2}
\bigg[A_3\\\nonumber&\times \bigg[2 A_3 (4 \xi -1) r^2 \left(8 \pi
\mathcal{B} r^2+6 \xi -1\right)+16 \left(2 \pi \mathcal{B} (4 \xi
-1) r^2+9 \xi ^2\right)\\\label{18}&-54 \xi +5\bigg]+16 \pi
\mathcal{B} (4 \xi-1)\bigg]\bigg].
\end{align}
Exploiting the property that $\tilde{P}_r\big|_{(r=\mathcal{G})}=0$,
we obtain the following explicit expression for the bag constant
\begin{equation}
\mathcal{B}=\frac{(6\xi-1)\left(3A_2 A_3^2\mathcal{G}^3+2 A_1
A_3\sqrt{A_3\mathcal{G}^2}+2A_2A_3\mathcal{G}\right)}{16\pi(4\xi-1)
\sqrt{A_3\mathcal{G}^2}\left(A_3\mathcal{G}^2+1\right){}^2\left[A_2
\mathcal{G}\sqrt{A_3\mathcal{G}^2}+2A_1\right]}.
\end{equation}

\section{Matching Conditions}

Matching conditions establish the criteria for seamlessly joining
the internal and external geometries at the surface of compact
objects. The selection of the external geometry is based on the
requirement that the fundamental characteristics (such as the
presence or absence of charge and whether the spacetime is static or
dynamic) of the outer and inner regions are consistent at the
spherical boundary. Given that the inner geometry, as described in
Eq.\eqref{10}, is not influenced by charge, the Schwarzschild metric
is the most appropriate choice for the external spacetime. The
Schwarzschild spacetime is preferred to the Schwarzschild-de Sitter
solution as the exterior geometry in our model because the field
equations do not include a cosmological constant $\Lambda$. The
Schwarzschild-de Sitter solution arises only in the presence of a
nonzero $\Lambda$, which is not part of our framework. Consequently,
using the Schwarzschild solution ensures consistency with the
theoretical assumptions and accurately reflects the absence of
$\Lambda$ in the governing equations. The outer Schwarzschild metric
is given by
\begin{equation}\label{19}
ds^2_+=\bigg(1-\frac{2\tilde{\mathcal{M}}}{r}\bigg)dt^2-\bigg(1-
\frac{2\tilde{\mathcal{M}}}{r}\bigg)^{-1}dr^2-r^2(d\theta^2
+\sin^2\theta d\phi^2),
\end{equation}
where $\tilde{\mathcal{M}}$ denotes the mass at the boundary
($r=\mathcal{G}$). We mention that in \cite{5bb}, the authors
obtained anisotropic stellar models by employing the Durgapal-Lake
ansatz in Rastall theory, using the Schwarzschild metric to denote
the outer geometry in the matching conditions.

The following constraints arise from the requirement that the first
fundamental form remains continuous at the surface
\begin{align}\label{20}
g_{tt}:&~~\big(A_1+\frac{1}{2}A_2\mathcal{G}\sqrt{A_3\mathcal{G}^2}\big)^2
=1-\frac{2\tilde{\mathcal{M}}}{\mathcal{G}},\\\label{21}g_{rr}:&~~\frac{1}
{1+A_3\mathcal{G}^2}=1-\frac{2\tilde{\mathcal{M}}}{\mathcal{G}},\\\label{22}
g_{tt,r}:&~~A_2 \left(2 A_1 \sqrt{A_3 \mathcal{G}^2}+A_2 A_3
\mathcal{G}^3\right)=\frac{2\tilde{\mathcal{M}}}{\mathcal{G}^2}.
\end{align}
From Eqs.\eqref{20}-\eqref{22} above, the parameters $(A_1,A_2,A_3)$
are obtained as
\begin{align}\label{23}
A_1&=\frac{\sqrt{\frac{\tilde{\mathcal{M}}}{\mathcal{G}-2
\tilde{\mathcal{M}}}} (2 \mathcal{G}-5 \tilde{\mathcal{M}})}{2
\sqrt{\tilde{\mathcal{M}}\mathcal{G}}},\\\label{24}A_2&=\frac{\sqrt{\tilde{\mathcal{M}}}}
{\sqrt{2}\mathcal{G}^{3/2}},\\\label{25}A_3&=-\frac{2\tilde{\mathcal{M}}}
{\mathcal{G}^2(2\tilde{\mathcal{M}}-\mathcal{G})}.
\end{align}
The constraints $A_1,A_2,A_3$ have dimensions $L^0,L^{-1},L^{-2}$,
respectively, where $L$ denotes length. Thus $A_1$ has no units,
while the units of $A_2$ and $A_3$ are $m^{-1}$ and $m^{-2}$,
respectively, where $m$ denotes meters (the SI unit for length).
Observational data, including the measured masses and radii of five
distinct strange stars; \emph{SAX J} $\emph{1808.4-3658}$ \cite{40},
\emph{Her X-1} \cite{41}, \emph{PSR J} $\emph{038-0842}$ \cite{41},
\emph{SMC X-1} \cite{42}, and \emph{LMC X-4} \cite{42}, have been
analyzed. \textbf{Table 1} presents these details, along with the
ratio of mass to radius for each star, expressed as a dimensionless
parameter. Our results indicate that the calculated ratios remain
within the upper bound \( \frac{\tilde{\mathcal{M}}}{2\mathcal{G}} <
\frac{2}{9} \), in accordance with Buchdahl's criterion \cite{43}.
Moreover, \textbf{Table 2} provides the corresponding values for the
constraints $(A_1,A_2,A_3)$ in the Finch-Skea metric, derived from
the data in \textbf{Table 1}. Finally, \textbf{Table 3} presents
calculated values of the bag constants for $\xi=0.3,0.5$, for each
of the quark candidates considered.
\begin{table}
\centering \caption{Mass and radius of some strange stars}
\vspace{0.2cm}
\begin{tabular}{|c|c|c|c|}
\hline Star & $\mathcal{G}$ (km) & $\tilde{\mathcal{M}}$
(km) & $\frac{\tilde{\mathcal{M}}}{\mathcal{G}}$\\
\hline
\emph{SAX J 1808.4-3658} & 7.07 & 2.124 & 0.300424 \\
\hline
\emph{Her X-1} & 8.1 & 1.25375 & 0.154784 \\
\hline
\emph{LMC X-4} & 8.831 & 1.90275 & 0.215463 \\
\hline
\emph{PSR J038-0842} & 10.06 & 3.0975 & 0.307903 \\
\hline
\emph{SMC X-1} & 9.34 & 1.534 & 0.16424 \\
\hline
\end{tabular}
\end{table}
\begin{table}
\centering \caption{Values of parameters$(A_1,A_2,A_3)$}
\vspace{0.2cm}
\begin{tabular}{|c|c|c|c|}
\hline Star & $A_1$ & $A_2(m^{-1})$ & $A_3(m^{-2})$\\
\hline
\emph{SAX J 1808.4-3658} & 3.9463 $\times 10^{-1}$ & 5.48007 $\times 10^{-2}$& 3.00643 $\times 10^{-2}$\\
\hline
\emph{Her X-1} & 7.37986 $\times 10^{-1}$ & 3.43333 $\times 10^{-2}$& 0.682713 $\times 10^{-2}$\\
\hline
\emph{LMC X-4} & 6.11888 $\times 10^{-1}$ & 3.71547 $\times 10^{-2}$& 0.969829 $\times 10^{-2}$\\
\hline
\emph{PSR J038-0842} & 3.721 $\times 10^{-1}$ & 3.89894 $\times 10^{-2}$& 1.58099 $\times 10^{-2}$\\
\hline
\emph{SMC X-1} & 7.19472 $\times 10^{-1}$ & 3.06712 $\times 10^{-2}$& 0.560166 $\times 10^{-2}$\\
\hline
\end{tabular}
\end{table}
\begin{table}
\centering \caption{Values of Bag constant $\mathcal{B}$ for
$\xi=0.3,0.5$}
\vspace{0.2cm}
\begin{tabular}{|c|c|c|}
\hline Star Models & $\mathcal{B}\big|_{\xi=0.3}(km^{-2})$
& $\mathcal{B}\big|_{\xi=0.5}(km^{-2})$\\
\hline
\emph{PSR J038-0842} & 4.28051 $\times 10^{-4}$ &  2.14026 $\times 10^{-4}$\\
\hline
\emph{SAX J 1808.4-3658} & 8.59907 $\times 10^{-4}$ &  4.29953 $\times 10^{-4}$\\
\hline
\emph{SMC X-1} & 3.50867 $\times 10^{-4}$ &  1.75434 $\times 10^{-4}$\\
\hline
\emph{LMC X-4} & 4.69899 $\times 10^{-4}$ &  2.34949 $\times 10^{-4}$\\
\hline
\emph{Her X-1} & 4.46748 $\times 10^{-4}$ &  2.23374 $\times 10^{-4}$\\
\hline
\end{tabular}
\end{table}

The value of this constant is not entirely arbitrary, as it is
generally accepted to fall within the range $( 57 \leq \mathcal{B}
\leq 92$) $MeV/fm^3$, as reported by Fiorella Burgio and Fantina
\cite{38}. In their study \cite{38a}, the authors determined
possible ranges for $\mathcal{B}$ by analyzing $20$ compact star
candidates without assuming a specific value a priori. Their
findings suggest that the Bag constant lies within the range of
$41.58 MeV/fm^3$ to $333.41 MeV/fm^3$, depending on the mass and
radius of the observed stars.

\section{Physical Analysis}

This sections examines diverse structural attributes of strange
stars by employing an anisotropic model within the context of
Rastall gravity. Using the data in \textbf{Table 1}, we examine the
graphical trends of several matter variables. In this analysis, we
evaluate various aspects of quark stars, including the feasibility
of their metric potentials, energy density, and anisotropic
pressure. We also examine the energy bounds, compactness, and
surface redshift. Additionally, we evaluate their stability. A
consistent solution ensures that the metric components are free from
singularities, exhibit a monotonically increasing pattern, and
maintain positive values throughout. As shown by Eq.\eqref{15}, the
metric coefficients are exclusively determined by the Finch-Skea
constants, with their computed values presented in \textbf{Table 2}.
Figure \textbf{1} illustrates graphical behavior of the metric
functions, thereby validating the physical accuracy of the proposed
solution. Where applicable in all the considered plots, we have used
the calculated values of the bag constant $\mathcal{B}$ presented in
\textbf{Table 3} while $\xi=0.3,0.5$ are denoted by thick and dashed
lines, respectively. In what follows, the colors black, red, green,
brown and blue, denote the stars \emph{SMC} $\emph{X-1}$, \emph{LMC}
$\emph{X-4}$, \emph{PSR} $\emph{J038-0842}$, \emph{SAX J}
$\emph{1808.4-3658}$, and \emph{Her} $\emph{X-1}$, respectively.

While the $\mathbb{MIT}$ bag model $\mathbb{EOS}$ is specifically
designed for strange quark matter, several compact objects,
including pulsars, have been proposed as strange star candidates
based on observational constraints. Notably, stars such as PSR
$J038-0842$ and PSR $J0740+6620$ have been investigated in the
literature under the assumption that they could contain deconfined
quark matter at extreme densities \cite{40,41}. The selection of
stars in our study follows similar reasoning, as their measured
mass-radius relationships and surface properties are compatible with
strange star models. Moreover, previous studies have successfully
applied the $\mathbb{MIT}$ bag model to pulsars, reinforcing its
applicability to such objects \cite{22,23}.
\begin{figure}\center
\epsfig{file=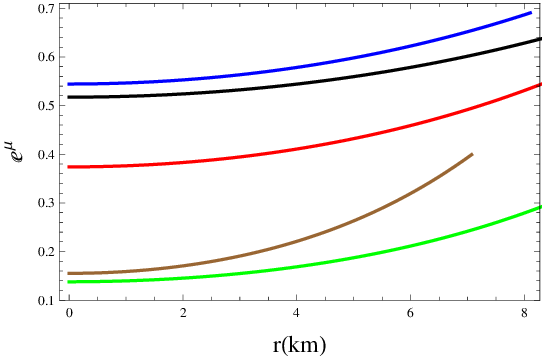,width=0.475\linewidth}
\epsfig{file=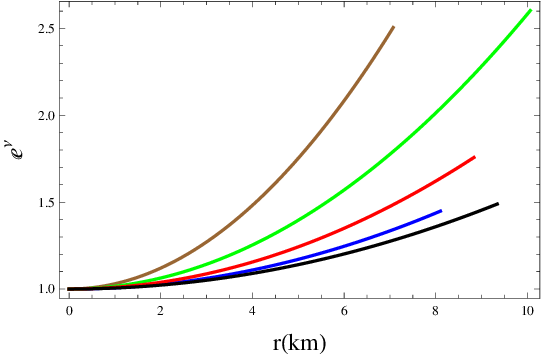,width=0.475\linewidth}\caption{$g_{tt}$ and
$g_{rr}$.}
\end{figure}

\subsection{State Variables}

The matter in a proper fluid configuration tends to be concentrated
at the center. Consequently, a solution is deemed physically valid
only if the properties of the matter, such as density and pressure,
are at their highest at the core and gradually decrease towards the
outer boundary. Since the fluid in question exhibits anisotropy, our
analysis focuses on three key aspects: the $\tilde{\rho}$,
$\tilde{P}_r$ and $\tilde{P}_t$. $\tilde{P}_r$ is expected to
approach zero at the boundary. As shown in Figure \textbf{2}, these
matter properties meet the necessary criteria. For comparison sake,
we also investigate the behavior of matter variables for the
vanishing Rastall parameter (GR case) in Figure \textbf{3}. It is
observed that in this case the matter variables also display
acceptable behavior as in the case of non-vanishing Rastall
parameter discussed above. However, the density and radial pressure
are higher at the core as compared to the non-vanishing case.
Additionally, we assess the regularity of these matter variables
using the conditions $\frac{d\tilde{\rho}}{dr}<0,
\frac{d\tilde{P}_r}{dr}<0, \frac{d\tilde{P}_t}{dr}<0$. Figure
\textbf{4} demonstrates that the matter properties conform to these
regularity conditions, thereby implying a highly compact anisotropic
matter distribution within this theoretical framework. Additionally,
we observe that a lower Rastall parameter induces a denser core as
well as a higher radial pressure in the core.
\begin{figure}\center
\epsfig{file=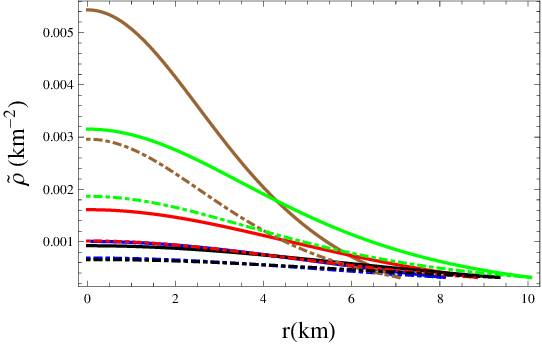,width=0.475\linewidth}
\epsfig{file=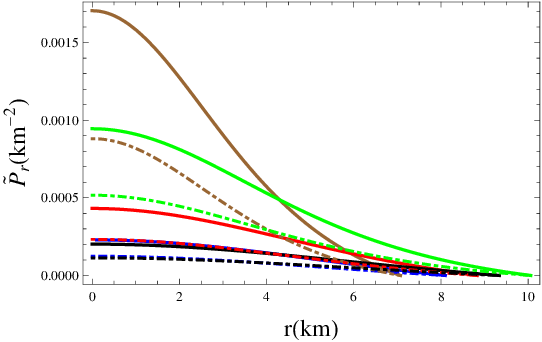,width=0.475\linewidth}
\epsfig{file=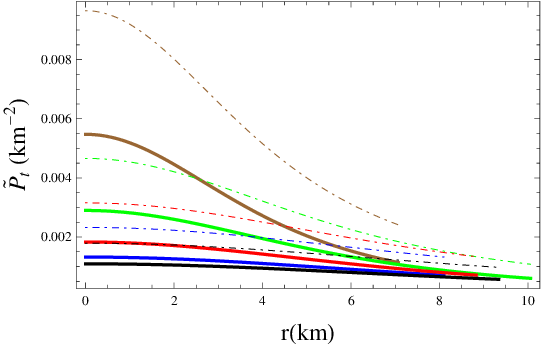,width=0.475\linewidth}\caption{Graphs of matter
variables against $r$ for $\xi=0.3$ (solid), $0.5$ (dashed).}
\end{figure}
\begin{figure}\center
\epsfig{file=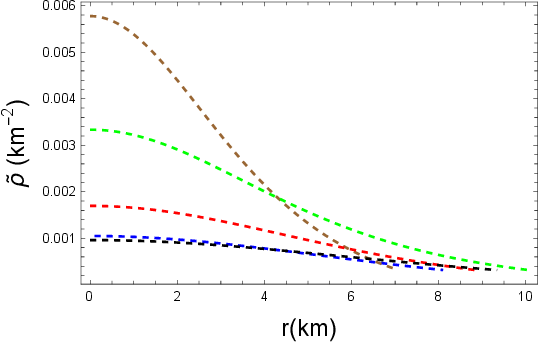,width=0.475\linewidth}
\epsfig{file=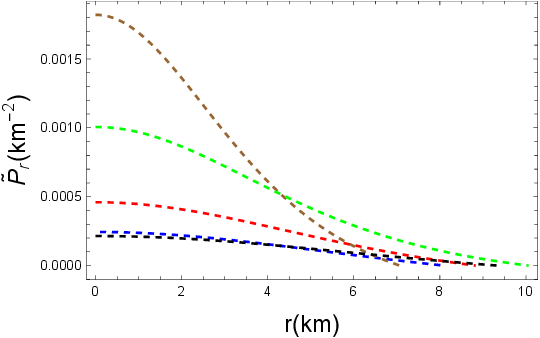,width=0.475\linewidth}
\epsfig{file=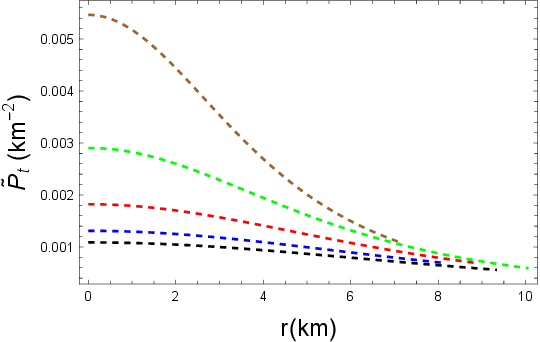,width=0.475\linewidth}\caption{Graphs of matter
variables against $r$ for $\xi=0$.}
\end{figure}
\begin{figure}\center
\epsfig{file=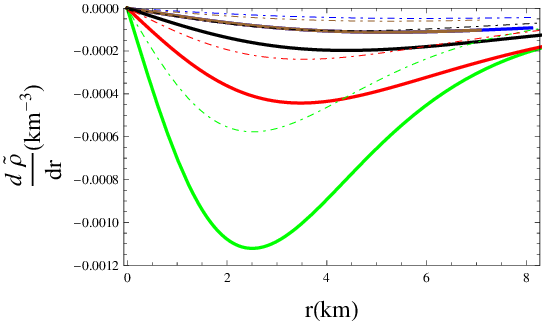,width=0.475\linewidth}
\epsfig{file=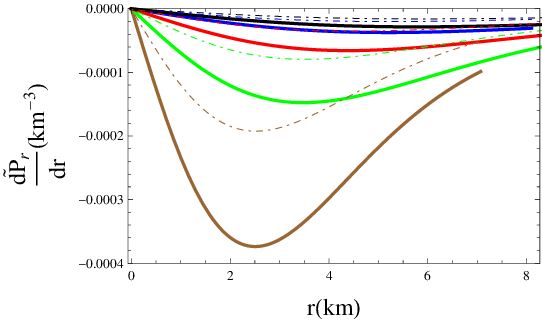,width=0.475\linewidth}
\epsfig{file=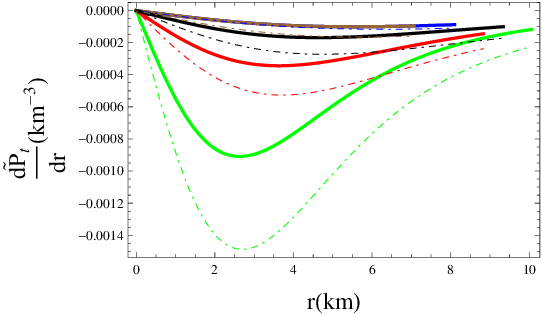,width=0.475\linewidth}\caption{Graphs of
$\frac{d\tilde{\rho}}{dr},\frac{d\tilde{P}_r}{dr},$ and
$\frac{d\tilde{P}_t}{dr}$ against $r$ for $\xi=0.3$ (solid), $0.5$
(dashed).}
\end{figure}

\subsection{Anisotropic Pressure}

The fluid anisotropy $(\tilde{\Delta})$, which is due to the
directional variation of the fluid pressure, is defined as
$\tilde{\Delta}=\tilde{P}_t-\tilde{P}_r$. Using Eqs.\eqref{17} and
\eqref{18}, this parameter turns out to
\begin{align}\nonumber
\tilde{\Delta}&=\frac{1}{8 \pi \left(A_3 r^3+r\right)^2 \left(4 A
\left(A_1+A_2 r \sqrt{A_3 r^2}\right)+A_2^2 A_3 r^4\right)}\bigg[r
\bigg[4 A_1^2 r \bigg[A_3 \bigg[r^2
\bigg[A_3\\\nonumber&\times\bigg(16 \pi \mathcal{B} r^2+6 \xi
-1\bigg)+32 \pi \mathcal{B}\bigg]+18 \xi -3\bigg]+16 \pi
\mathcal{B}\bigg]+4 A_1A_2 \sqrt{A_3 r^2}\\\nonumber&\times
\left(r^2 \left(A_3 r^2+1\right) \left(A_3 \left(16 \pi\mathcal{B}
r^2+6 \xi -1\right)+16 \pi\mathcal{B}\right)-18 \xi +3\right)+A_2^2
A_3r^3\\\nonumber&\times \bigg[(r^2 \left(A_3 \left(r^2 \left(A_3
\left(16 \pi\mathcal{B}r^2+6 \xi -1\right)+32 \pi
\mathcal{B}\right)-6 \xi +1\right)+16 \pi\mathcal{B}\right)-36 \xi
\\\label{26}&+6\bigg]\bigg]\bigg].
\end{align}
Anisotropic pressure is characterized by differences between the
tangential and radial components. When the tangential pressure
exceeds the radial pressure $(\tilde{P}_t>\tilde{P}_r)$, it
indicates an outward force, whereas the reverse scenario
$(\tilde{P}_r > \tilde{P}_t)$ signifies an inward force. The
presence of positive anisotropic pressure generates an
outward-directed force that opposes the inward gravitational pull in
stellar bodies, helping to maintain equilibrium. Figure \textbf{5}
illustrates the anisotropic pressure distribution for the chosen
quark star models. The parameter $\tilde{\Delta}$ decreases
monotonically towards the surface while maintaining a positive value
throughout, indicating a repulsive force that plays a role in the
structural development of massive stellar objects.
\begin{figure}\center
\epsfig{file=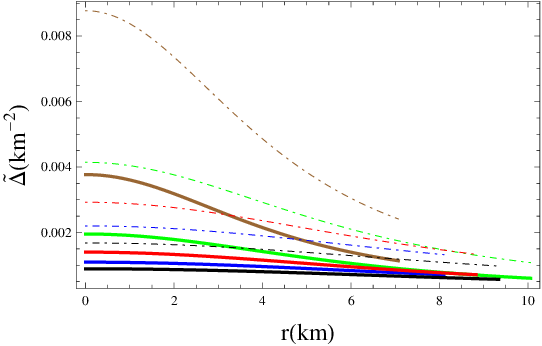,width=0.5\linewidth}\caption{$\tilde{\Delta}$
versus $r$ for $\xi=0.3$ (solid), $0.5$ (dashed).}
\end{figure}
Anisotropic pressure can change the equilibrium configuration of
quark stars. When the tangential pressure exceeds the radial
pressure, as is the case in this work, an outward directed pressure
is produced. The outward pressure that results when the tangential
pressure supersedes the radial pressure could counterbalance the
gravitational attraction, resulting in a more expanded radius at
higher central densities than one would consider for an isotropic
configuration. Such situation could explain the absence of the
radius reducing trend as the maximum mass is approached, which is
why we see an increasing radius in the mass-radius plots made.
Anisotropic pressures may also allow the maintenance of higher mass
stars by means of the readjustment of internal pressures. This
readjustment makes it possible to achieve greater central pressures
in the star without the normal instability that usually occurs in
isotropic models and causes the fall in radius close to the mass
limit. This could perhaps suggest that the central density rises
monotonically with the dimensions of the star even near the maximal
mass limit.

\subsection{Mass, Compactness, and Surface Redshift}
\begin{figure}\center
\epsfig{file=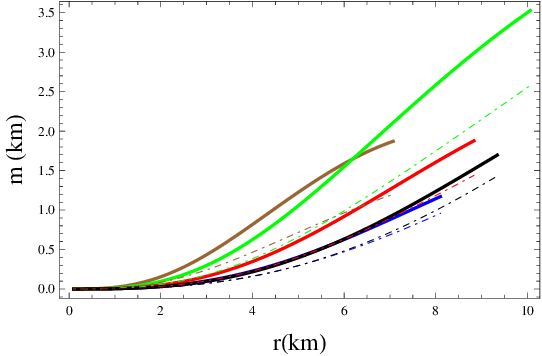,width=0.475\linewidth}
\epsfig{file=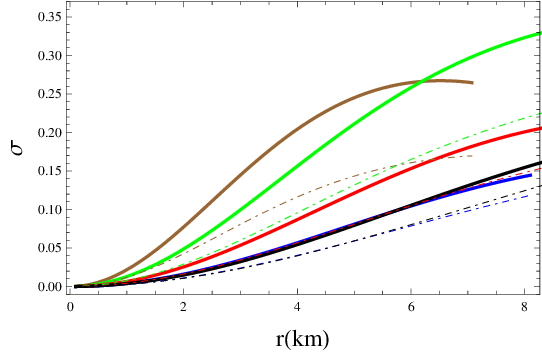,width=0.475\linewidth}
\epsfig{file=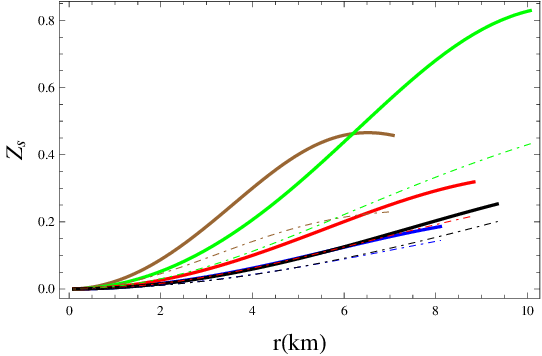,width=0.475\linewidth}\caption{Graphs of
$m(r)$, $\sigma(r)$, and $Z_s(r)$ against $r$ for $\xi=0.3$ (solid),
$0.5$ (dashed).}
\end{figure}

The mass of a spherical object can be calculated based on the energy
density using the equation \cite{42a}
\begin{equation}\label{27}
m(r)=\int_0^\mathcal{G}4\pi\tilde{\rho}r^2 dr,
\end{equation}
where $\tilde{\rho}$ is defined in Eq.\eqref{16}. Figure \textbf{6}
exhibits the vanishing of this function at the core and also shows
its monotonic increase towards the surface. Additionally, for the
Rastall parameter value $\xi = 0.3$, the model offers a reasonable
estimate for the mass of quark stars. We also study the compactness,
given by $\sigma(r)-\frac{m(r)}{r}=0$. Compactness quantifies the
degree to which an object's mass is confined within a given radius,
playing a vital role in assessing the intensity of the gravitational
field at the stellar surface. The change in the wavelength of
electromagnetic radiation of a dense astrophysical body is
characterized by the gravitational redshift, represented as $(
Z_s(r)$). Owing to the intense field near the surface, the energy of
the emitted radiation diminishes, leading to an elongation of its
wavelength, commonly referred to as redshift. Photons originating
from deeper within the core must travel through denser regions,
losing energy due to scattering. In contrast, photons emitted near
the surface encounter less dense matter, leading to less scattering
and reduced energy loss. To ensure a stable configuration, the
conditions $\sigma < \frac{4}{9}$ \cite{43} and $Z_s \leq 5.2$
\cite{44} must be satisfied. The plots of compactness and surface
redshift in Figure \textbf{6} confirm that our model adheres to
these stability limits.

\subsection{Energy Conditions}

In astrophysics, the type of matter present within a body is often
confirmed through specific constraints called energy constraints.
These constraints are essential for comprehending the
characteristics and behavior of celestial bodies. These conditions
allow us to differentiate between ordinary and exotic matter within
a given geometry. The satisfaction of these conditions, which depend
on various physical quantities like $\tilde{P}$ and $\tilde{\rho}$,
confirms the presence of normal matter within a compact star.
Furthermore, these limits are instrumental in evaluating the
practicality of proposed models within various gravitational
theories. To ensure that a particular geometric configuration
accommodates ordinary matter, it is essential that its corresponding
physical parameters adhere to specific criteria. These criteria can
be categorized as follows
\begin{figure}\center
\epsfig{file=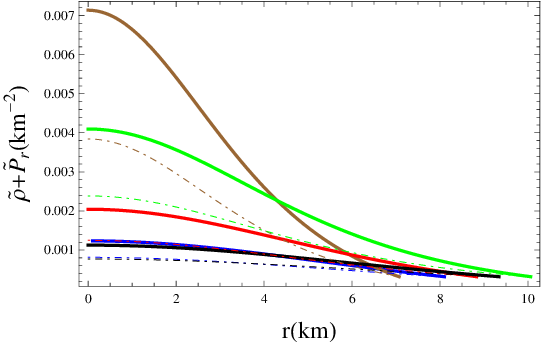,width=0.475\linewidth}
\epsfig{file=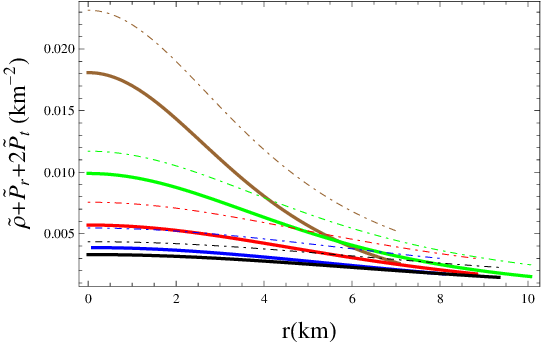,width=0.475\linewidth}
\epsfig{file=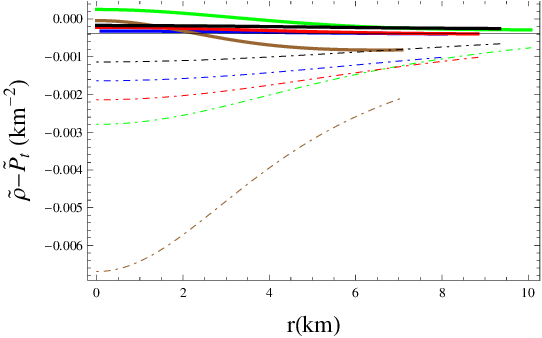,width=0.475\linewidth}
\epsfig{file=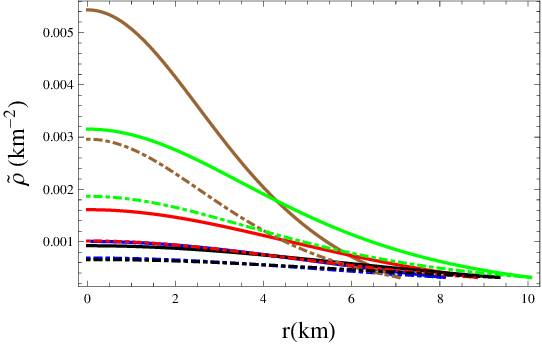,width=0.475\linewidth}
\epsfig{file=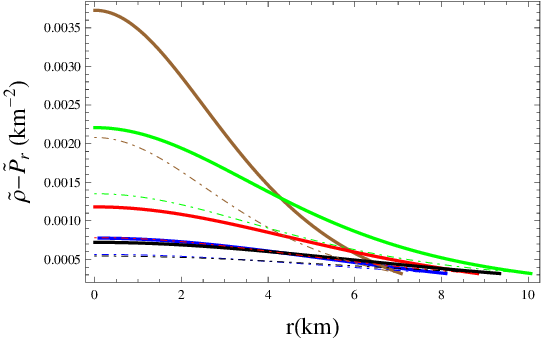,width=0.475\linewidth}
\epsfig{file=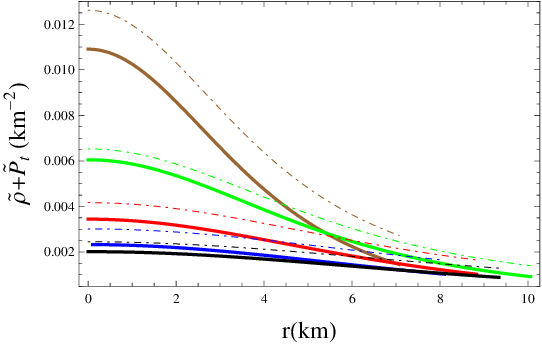,width=0.475\linewidth} \caption{Energy conditions
for $\xi=0.3$ (solid), $0.5$ (dashed).}
\end{figure}
\begin{itemize}
\item Dominant Energy Conditions\\
$\tilde{\rho}\geq |\tilde{P_r}|,\quad\tilde{\rho}\geq
|\tilde{P_t}|.$
\item Strong Energy Conditions\\
$\tilde{\rho}\geq-\tilde{P_r},\quad\tilde{\rho}\geq-\tilde{P_t},
\quad\tilde{\rho}+\tilde{P_r}\geq-2\tilde{P_t}.$
\item Null Energy Conditions\\
$\tilde{\rho}\geq-\tilde{P_r},\quad\tilde{\rho}\geq-\tilde{P_t}.$
\item Weak Energy Conditions\\
$\tilde{\rho}\geq 0,\quad\tilde{\rho}\geq-\tilde{P_r},
\quad\tilde{\rho}\geq-\tilde{P_t}.$
\end{itemize}
Figure \textbf{7} illustrates the conditions under which a violation
of dominant energy conditions is evident. This observed violation
suggests the presence of exotic substances within the internal
composition of the quark candidates.

Retaining the GR forms of energy conditions allows the results to be
directly compared to observational constraints and to solutions in
GR, which is essential for establishing the validity of modified
gravity theories. Deviations from GR predictions can then be
attributed to differences in field equations rather than
reformulated energy conditions. While modified definitions of energy
conditions in Rastall gravity have been explored in some studies
(e.g., \cite{23b,40}), adopting the classical GR forms remains a
practical and widely accepted approach \cite{5bb,21,22,23}. In our
work, we have chosen to adhere to the classical GR definitions of
the energy conditions to maintain consistency with the broader
literature and to provide clear, direct comparisons with
observational data.

\subsection{Stability}

The stability of compact stars is of significant interest in
astrophysics, as it aids in developing physically viable models for
such objects. Massive celestial bodies that exhibit stable behavior
despite external disturbances are particularly fascinating, making
the study of their structural stability crucial. In the context of
Rastall theory, we employ three distinct approaches to examine the
stability of these compact objects.
\begin{figure}\center
\epsfig{file=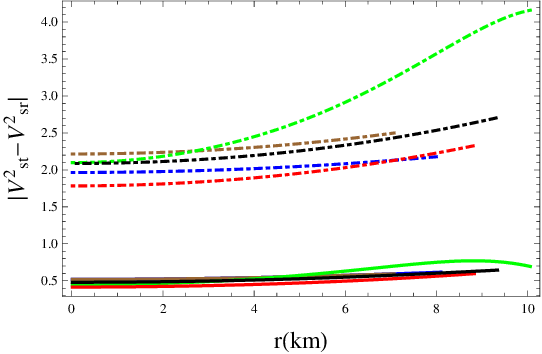,width=0.475\linewidth}
\epsfig{file=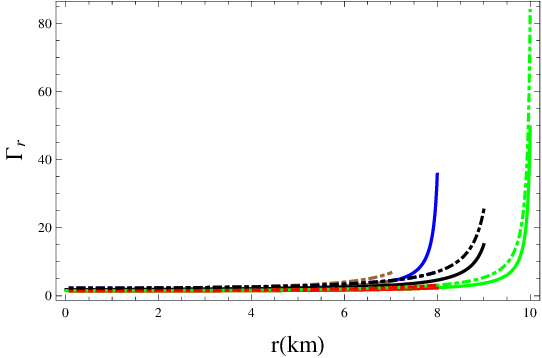,width=0.475\linewidth}
\epsfig{file=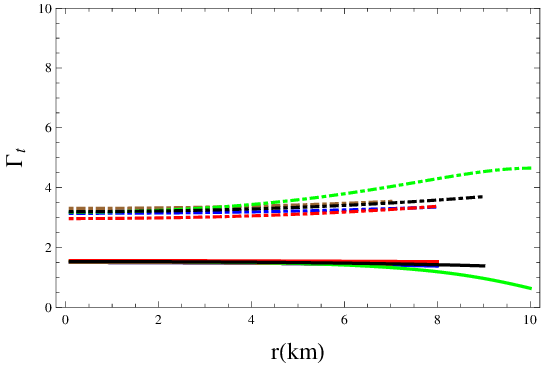,width=0.475\linewidth}\caption{$|V_{st}^2-V_{sr}^2|$,
$\Gamma_r$, $\Gamma_t$ for $\xi=0.3$ (solid), $0.5$ (dashed).}
\end{figure}

To start with, we employ the cracking approach proposed by Herrera
\cite{12}. In this concept, stability is ensured if the condition $0
\leq |V_{st}^2 - V_{sr}^2| \leq 1$ is met, where $V_{st}^2 =
\frac{d\tilde{P}_t}{d\tilde{\rho}}$ refers to the tangential sound
speed and $V_{sr}^2 = \frac{d\tilde{P}_r}{d\tilde{\rho}}$ denotes
the radial sound speed. We plot this property in Figure \textbf{8},
where it is illustrated that the configurations are stable for $\xi
= 0.3$ and unstable for $\xi = 0.5$. Additionally, we strengthen
this analysis using the adiabatic index method. With this approach,
a stable configuration is deduced if the adiabatic index remains
greater than $\frac{4}{3}$ \cite{46}. In the case of an anisotropic
configuration, this criterion is modified to $\Gamma_t >
\frac{4}{3}$ and $\Gamma_r
> \frac{4}{3}$, where $\Gamma_t$ and $\Gamma_r$ are the
tangential and radial adiabatic indices, respectively, given by
\begin{equation}\label{28}
\Gamma_r=\bigg(1+\frac{\tilde{\rho}}{\tilde{P}_r}\bigg)
\frac{d\tilde{P}_r}{d\tilde{\rho}},\quad
\Gamma_t=\bigg(1+\frac{\tilde{\rho}}{\tilde{P}_t}\bigg)
\frac{d\tilde{P}_t}{d\tilde{\rho}}.
\end{equation}

The corresponding stability conditions for the adiabatic indices are
also plotted in Figure \textbf{8}, which shows a stable regime for
the indices considered. Additionally, we examine the causality
conditions wherein $0\leq V_{st}^2,V_{sr}^2\leq 1$ for a stable
anisotropic configuration. This implies that both speed of sound
components must be contained in the closed interval $[0,1]$, for a
stable stellar configuration. We plot this in Figure \textbf{9},
where we observe a stable configuration only for Rastall parameter
$\xi=0.3$. The corresponding stability conditions for the adiabatic
indices are also plotted in Figure \textbf{8}, which shows a stable
regime for the indices considered. Additionally, we examine the
causality conditions wherein $0\leq V_{st}^2,V_{sr}^2\leq 1$ is
required for a stable anisotropic configuration. This implies that
both speed of sound components must be contained in the closed
interval $[0,1]$, for a stable stellar configuration. As depicted in
Figure \textbf{9}, this analysis reveals that stability is achieved
only for $(\xi = 0.3$).
\begin{figure}\center
\epsfig{file=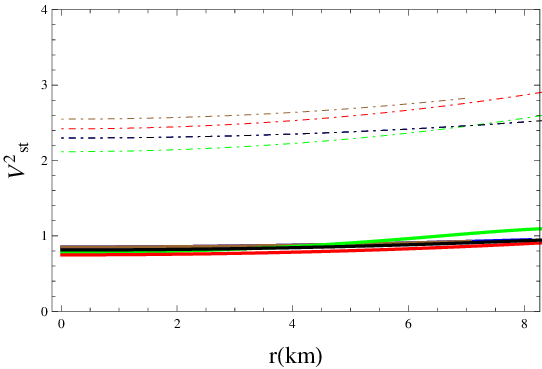,width=0.475\linewidth}
\epsfig{file=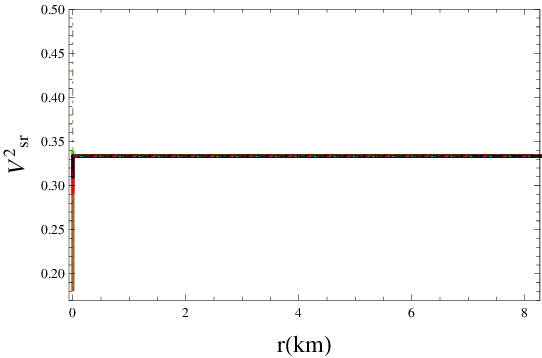,width=0.475\linewidth}\caption{$V_{st}^2$ and
$V_{sr}^2$ for $\xi=0.3$ (solid), $0.5$ (dashed).}
\end{figure}
\begin{figure}\center
\epsfig{file=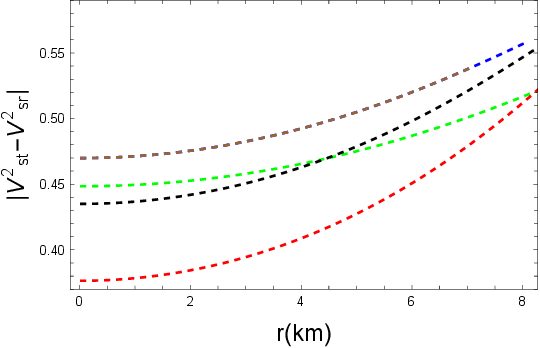,width=0.475\linewidth}
\epsfig{file=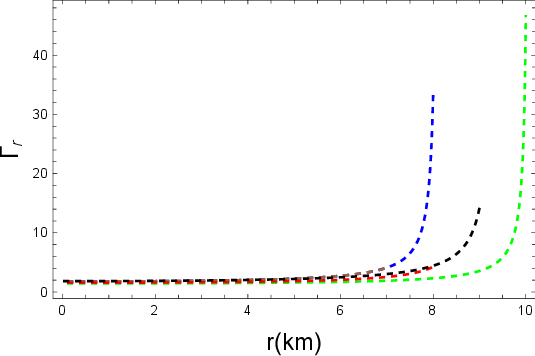,width=0.475\linewidth}
\epsfig{file=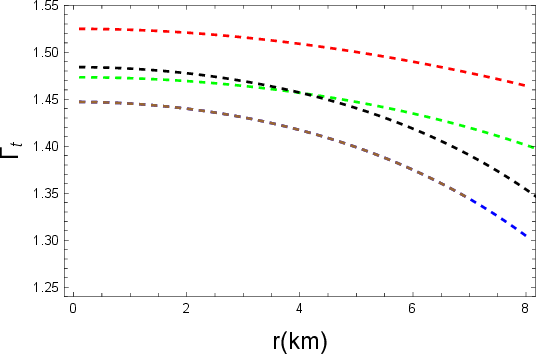,width=0.475\linewidth}
\caption{$|V_{st}^2-V_{sr}^2|$, $\Gamma_r$, and $\Gamma_t$ for
$\xi=0$.}
\end{figure}
\begin{figure}\center
\epsfig{file=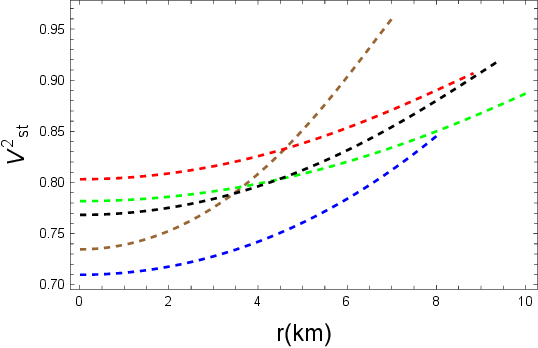,width=0.475\linewidth}
\epsfig{file=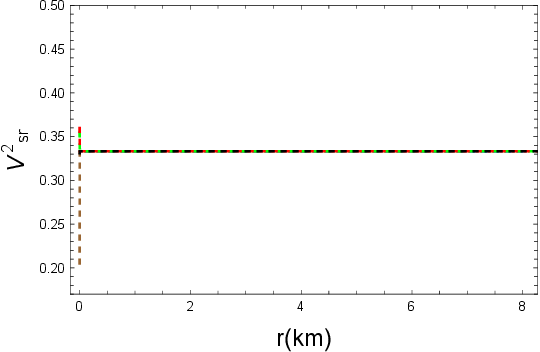,width=0.475\linewidth}\caption{$V_{st}^2$ and
$V_{sr}^2$ for $\xi=0$.}
\end{figure}

Finally, we also investigate the stability of the models when
considered with the vanishing Rastall parameter. The cracking
condition as well as the adiabatic indices shown in Figure
\textbf{10}, depict a stable model with respect to the vanishing
Rastall parameter. This result is further verified by the causality
conditions (Figure \textbf{11}).

\section{Conclusions}

This work constructs a theoretical model describing strange
anisotropic compact stars within the context of Rastall gravity. To
analyze the internal structure of the five specific stellar objects,
\emph{SAX J 1808.4-3658}, \emph{LMC X-4}, \emph{PSR J 038-0842},
\emph{Her X-1}, and \emph{SMC X-1}, we incorporate the
$\mathbb{EOS}$ derived from the $\mathbb{MIT}$ bag model alongside
the modified field equations of Rastall theory. The stellar geometry
is characterized by the Finch-Skea metric, which depends on three
undetermined parameters, $((A_1, A_2, A_3))$. These parameters are
expressed in terms of the observed stellar masses and radii by
applying the appropriate matching conditions dictated by Rastall's
framework. Using available astrophysical measurements, the
mass-to-radius ratios for these quark star candidates are evaluated
and summarized in \textbf{Table 1}. Further, \textbf{Table 2}
presents the computed values of the Finch-Skea parameters $( (A_1,
A_2, A_3)$), while \textbf{Table 3} provides the bag constant $(
\mathcal{B}$) corresponding to each stellar configuration,
considering various choices for the Rastall parameter.

In our study, we have focused on the values $\xi=0.3$ and $\xi=0.5$
to provide a detailed analysis of some physical features such as the
stability and physical viability of obtained stellar configurations.
While the Rastall parameter is not intrinsically constrained, apart
from certain specific values such as $\xi=\frac{1}{4}$ and
$\xi=\frac{1}{6}$, it is indeed impractical to explore the effects
of an infinite range of parameter values within a single study. The
selected values were chosen to highlight contrasting stability
behaviors and demonstrate the sensitivity of stellar stability to
changes in $\xi$. However, we acknowledge that investigating a
broader spectrum of $\xi$ could yield additional insights into the
parameter role in stellar configurations. Such an endeavor would
require a dedicated study, which we consider a valuable direction
for future research. Additionally, we have also investigated the
behavior of matter variables as well as the stability of the model,
with regards to the vanishing Rastall parameter $(\xi=0)$. We have
found that, the model is stable in addition to the acceptable
behavior of matter variables.

Graphical analyses are conducted to examine various physical
properties of the quark star candidates. The matter variables adhere
to the known characteristics of compact objects, specifically to the
maximality condition. As one moves towards the surface, both density
and pressures exhibit a monotonic decrease. Additionally, a positive
anisotropy is observed throughout. The measured redshift and
compactness values are within the expected bounds, and the mass
function suggests that our model provides a good approximation for
the mass and radius of strange stars when the Rastall parameter
$\xi=0.3$ is considered. However, the energy conditions are not
fully satisfied due to the violation of the dominant energy
condition, which indicates unusual matter in the interior of the
quark candidates. The stability of our model is assessed using the
Herrera cracking technique, adiabatic indices, and the causality
conditions, all of which indicate stable behavior for the model with
the Rastall parameter $\xi=0.3$. We observe that the stellar
configuration \emph{Her $X-1$} appears denser in this theory
compared to the $f(\mathcal{R},T,\mathcal{Q})$ framework \cite{23},
while \emph{SAX J $1808.4-3658$} is denser in this theory than in
the theories $f(\mathcal{R},T,\mathcal{Q})$ \cite{23} and
$f(\mathcal{R},T)$ \cite{47}.

Additionally, we compare our results to the results in \cite{5bb}.
Both studies explore the effects of Rastall gravity on compact stars
but differ in focus and methodology. Our study uses the Finch-Skea
metric to model anisotropic strange stars, employing the
$\mathbb{MIT}$ bag model to analyze their stability, compactness,
and surface redshift for specific Rastall parameter values
$\xi=0.3,0.5$. In contrast, the referenced work examines isotropic
stellar models using the Durgapal-Lake solutions, comparing
stability in Rastall gravity with GR and demonstrating that stars
are stable only in the Rastall framework. Unlike isotropic models in
Rastall gravity, such as Waseem and Naeem \cite{5bb}, our work
explores the impact of anisotropy on structural properties and
dynamical stability. Comparisons with recent studies, including
Pretel and Mota \cite{48} on hydrostatic equilibrium, El Hanafy
\cite{23b} on modeling PSR J0740+6620, and Nashed and El Hanafy
\cite{23aa} on stellar sizes, highlight the novel contributions of
our study in explicitly incorporating the MIT bag model EOS.
Additionally, Mota et al. \cite{49} examined anisotropic neutron
stars in Rastall-Rainbow gravity, and their findings align with our
results, reinforcing the significance of anisotropy in compact star
configurations. Finally, our results converge to those of GR when $\xi=0$.\\\\
\textbf{Data Availability Statement:} All data used are contained in
this paper and references therein.

\end{document}